\documentclass[12pt]{article}
\usepackage{blois,graphicx}

\begin{document}
\title{Average Transverse Momenta of Hadrons at LHC Energy 7 TeV vs. Masses and Heavy Neutral Hadron States}

\author{O.I. Piskounova}

\address{P.N.Lebedev Physical Institute of Russian Academy of Science, Leninski prosp. 53, 119991 Moscow, Russia}

\pagestyle{plain}

%
%\institute{Leninski prosp. 53, 119991 Moscow, Russia}
%
%\date{Received: date 26.08.2019/ Revised version: 26.08.2019}
\maketitle\abstract{
This paper examines the transverse momentum spectra of hadrons in the multi-particle production at LHC in the framework of Quark-Gluon String Model (QGSM). It discusses the dependence of average $p_t$ on the masses of mesons and baryons at the LHC energy $\sqrt{s}$=7 TeV. The QGSM description of experimental spectra of various hadrons led to the number of conclusions.

 I. The average transverse momenta of baryons and mesons are growing with the hadron mass and for beauty hadrons $<p_t>$ is almost equal to the mass. 

II. By the product of research, a regularity has been detected in the mass gaps between hadron generations. This hypothesis suggests some hidden symmetrical (neither-meson-nor-baryon) neutral hadron states with the masses: 0.251, 0.682, 1.85, 5.04, 13.7, 37.2, 101., 275., 748.... GeV, which are produced by geometrical progression with the mass factor of order $\delta$(lnM)=1.

III. The baryon-meson symmetry at low masses seems broken untill the mass of beauty hadrons, then the hidden states should be more and more stable with growing masses, so the suggested sequence of neutral hadronic states is a proper candidate for the Dark Matter, which, you know, contribute the valuable part to the mass of Universe .

The average transverse momenta have been extrapolated with the similar function, as for energy dependence of average baryon $p_t$, $<p_t> \propto M^{0.1}$.
}

\section{Introduction}
\label{intro}
The aim of this research was to analyze the transverse momentum spectra of hadrons from the modern collider experiments (ISR \cite{isr},STAR \cite{star}, ALICE \cite{alice}, and CMS \cite{cms}).
 
\begin{figure}[htpb]
  \centering
  \includegraphics[width=8.0cm, angle=0]{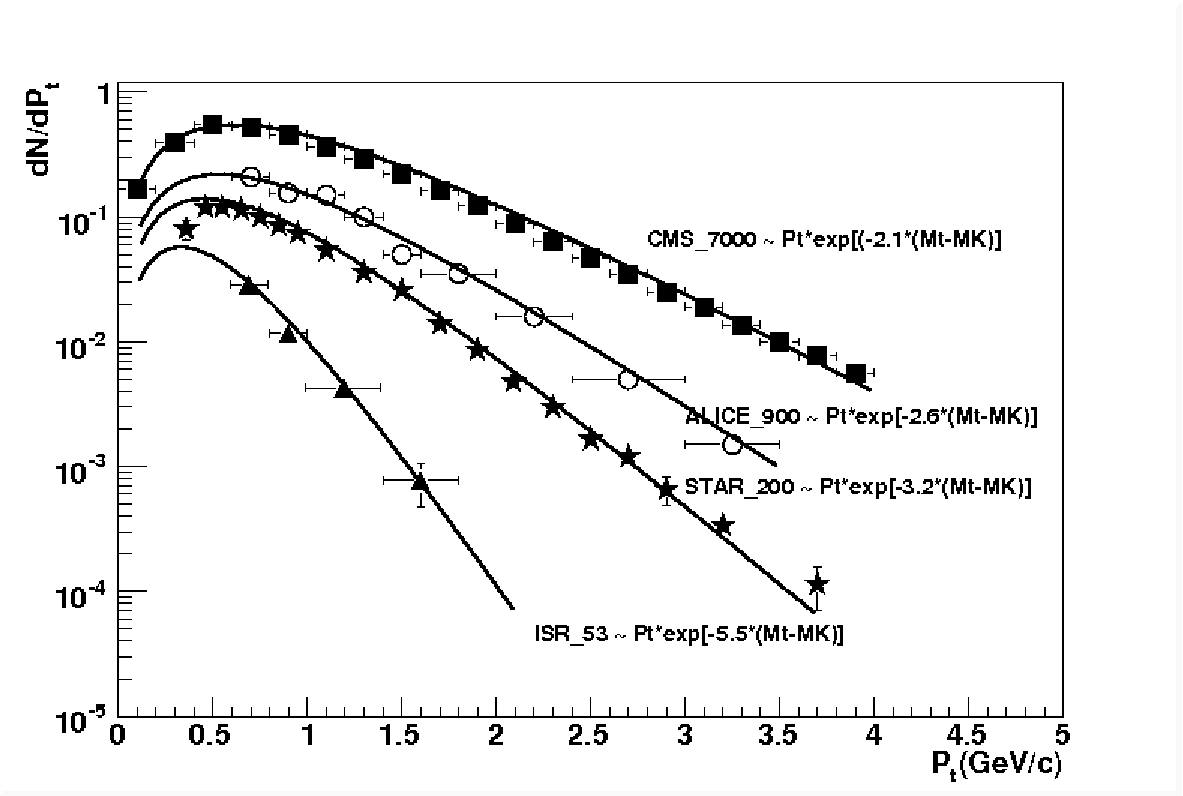}
  \caption{ Transverse momentum distributions of $\Lambda^0$ hyperons from colliders up to LHC. The data are from ISR \protect\cite{isr} $p-p$ at $\sqrt{s}= 53 GeV$ - black triangles, STAR \protect\cite{star} $p-p$ at $\sqrt{s}$=200 GeV - black stars; ALICE \protect\cite{alice} $p-p$ at $\sqrt{s}$=900 GeV - empty circles and CMS \protect\cite{cms} $\bar{p}-p$ at 7 TeV - black squares.}
  \label{spectracompiled}
\end{figure}

The figure~\ref{spectracompiled} presents the compilation of the data, $dN^(\Lambda^0)/dp_t$, in the region 0.1 GeV/c $< p_t <$ 5 GeV/c from the following experiments (ISR \cite{isr}, STAR \cite{star}, ALICE \cite{alice} and CMS \cite{cms}. It illustrates the changes in hyperon transverse distributions on the energy distance from ISR to LHC experiments. Since we do not need the absolute values of distributions in order to calculate the average $p_t$ , those are chosen arbitrarily. The range of low $p_t$ 0.3 GeV/c $< p_t <$ 4 GeV/c has the most impact on the value of average $p_t$. The figure clearly shows that average transverse momenta grow with energy.
In the previous publication \cite{avrgptvsenergy}, the growing values of baryon average $p_t$  are approximated with the power dependence, $<p_t> \propto s^{0.05}$.
The reason for this paper was an additional phenomenological research on the mass dependence of average transverse momenta of baryons. Since we have enough data for various hadrons, the spectra of mesons and baryons for all generations can be analyzed at the same LHC energy. 
This short paper  is devoted to the mass dependence of transverse momenta at the LHC energy $\sqrt{s}$= 7 TeV.

\section{Hadron Transverse Momentum Distributions in QGSM}
\label{sec:1}
  
\begin{figure}[htpb]
  \centering
  \includegraphics[width=8.0cm, angle=0]{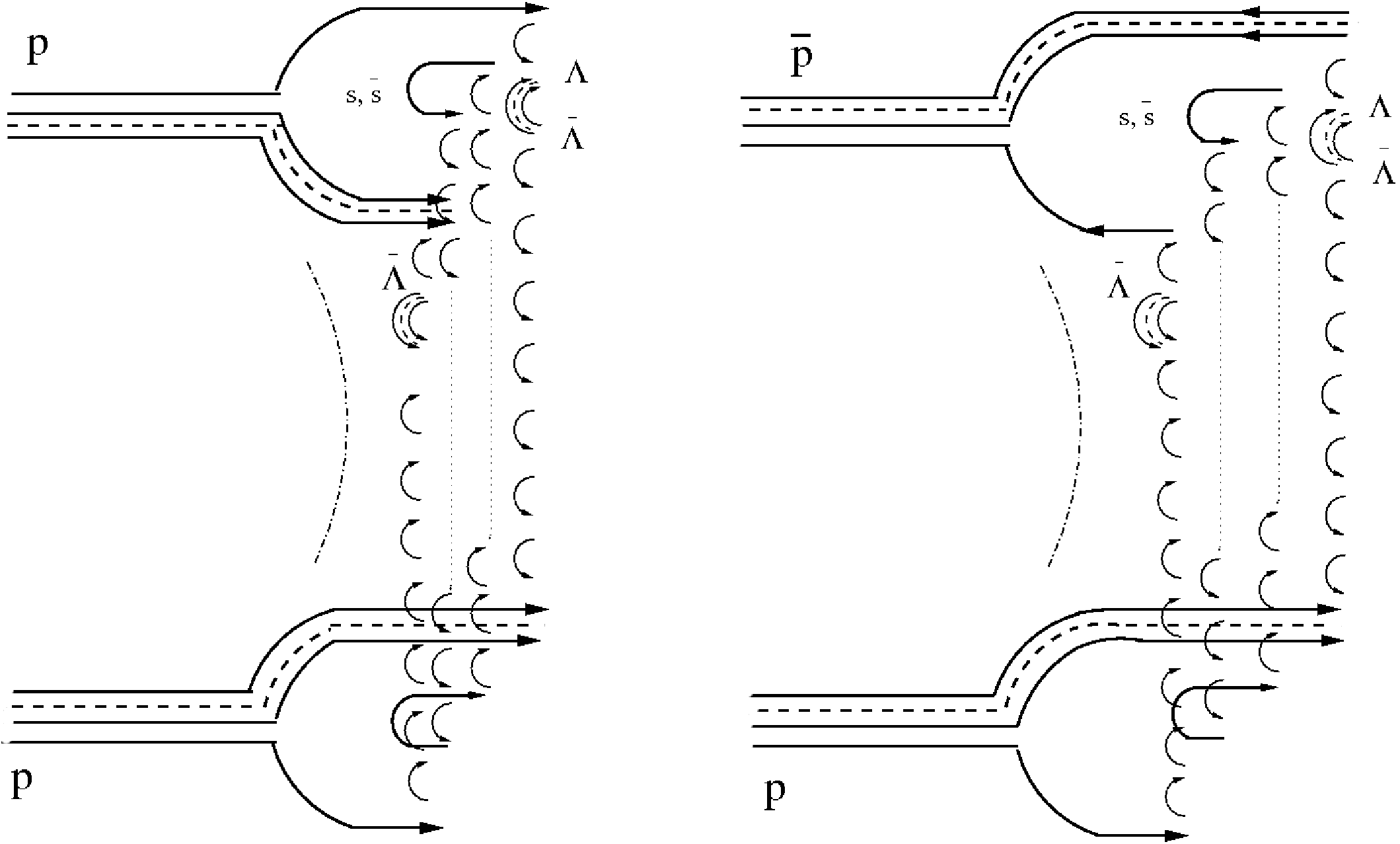}
  \caption{QGSM multiparticle production diagrams: the left is for proton-proton and the right is for antiproton-proton reactions.}
\label{antipomeron}	
\end{figure}

The Model of Quark-Gluon Strings \cite{qgsm,veselov,hyperon,piskounova} has been designed for description of rapidity distributions and hadron spectra in $x_F$ and in $p_t$ . The Model operates with pomeron diagrams, which help to calculate the spectra of produced hadrons. These spectra are presented as the convolutions of constituent quark structure functions with the quark-antiquark pair distributions at the fragmentation of cut pomeron cylinder sides into the certain hadrons (see the figure~\ref{antipomeron}). 

The early QGSM study \cite{veselov} on the hadron transverse momentum distributions has shown that the spectra of baryons in proton-proton collisions can be described with the following $p_t$-dependence:

\begin{equation}
E \frac{d^{3}\sigma^H}{dx_F d^{2}p_{t}}= \frac{d\sigma^{H}}{dx_F}*A_0*\exp[-B_0*(m_t-M)],
\end{equation}
where $M$ is the mass of produced hadron, $m_t$ = $\sqrt{p_t^2+M^2}$. The slope parameter, $B_0$, used to bring the dependence on $x_F$ in the previous research \cite{veselov}. The value of the slopes of baryon spectra for the data in central region of rapidity, Y=0, are constants and were estimated for many types of hadron spectra ($\pi$, K, p)  the collisions up to the energies of ISR experiment.
As it is seen above, the slopes of spectra, $B_0$, at the modern collider experiments depend on energy of collision, that gives the growing average transverse momenta. 

The comprehansive study for energy dependence of baryon average $p_t$ has been fulfilled in
the recent paper \cite{avrgptvsenergy}.
We can also deacribe the $p_t$ spectra for the production of other  hadrons  at the fixed energy of proton-proton collision and consider how they change with the growing mass of hadrons.

\section{Average Transverse Momenta vs. Mass of Hadrons}
\label{sec:3}

\begin{figure}[htpb]
  \centering
  \includegraphics[ width=8.0cm, angle=0]{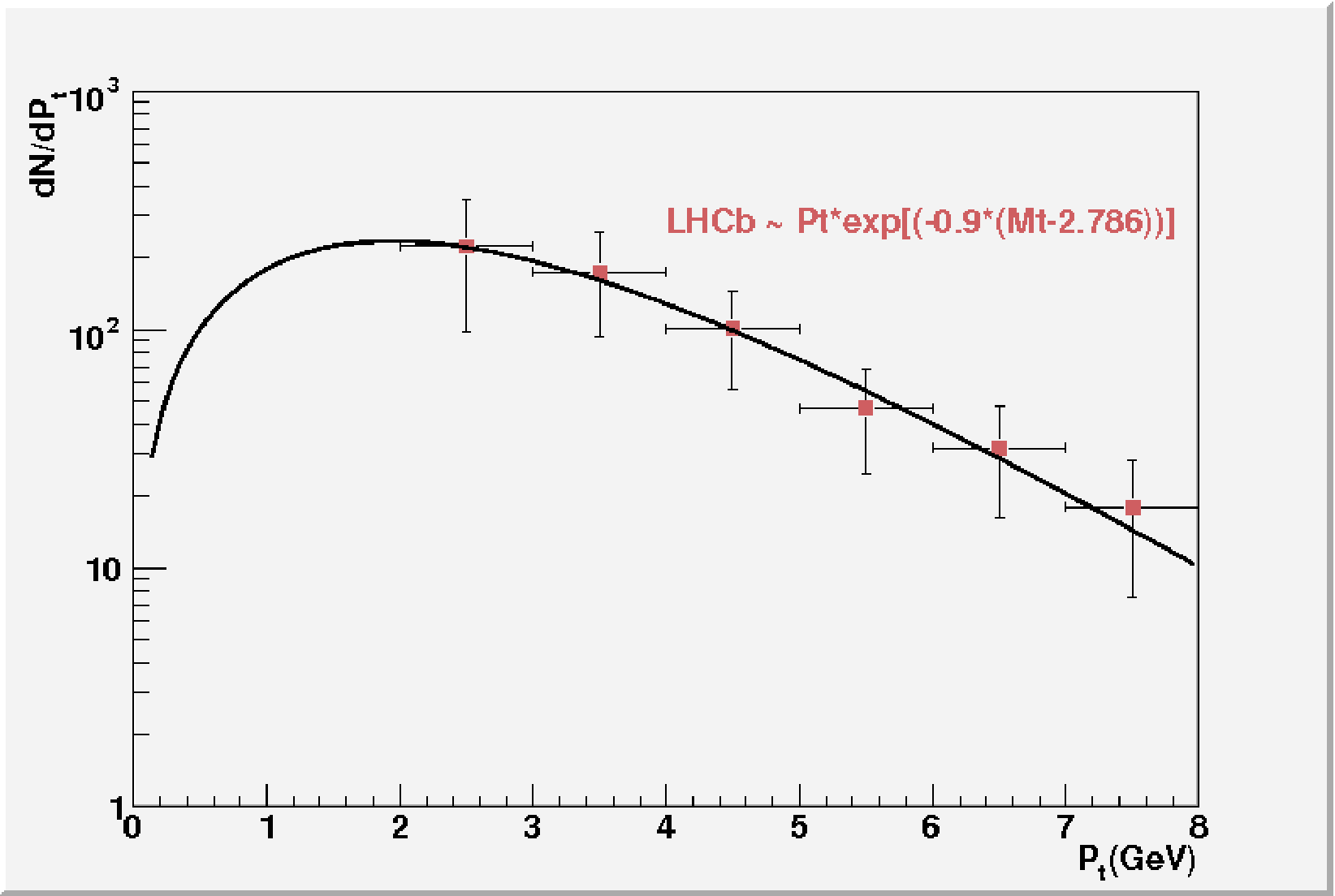}
  \caption{The QGSM fit of $\Lambda_c$ spectra at $\sqrt{s}$=7 TeV in LHCb experiment \protect\cite{lhcbmeson}.}
\label{HQspectra}	
\end{figure}

The previously published analysis of transverse momentum spectra of baryons from LHC experiments (ALICE, ATLAS, CMS)\cite{recent,bylinkin} was provided with only partial data on hadron spectra. In order to get full understanding of the average transverse momentum dependence on hadron mass, we supplement here the data on kaon, D-meson and B-meson spectra from LHC \cite{lhcbmeson} at 7 TeV. As an example, the figure~\ref{HQspectra} shows the $p_t$ distribution of $\Lambda_c$ production that  has been fitted with the same formula (1) as the $\Lambda^0$ hyperon spectra \cite{avrgptvsenergy}.
 
\begin{figure}[htpb]
  \centering
  \includegraphics[width=12.0cm, angle=0]{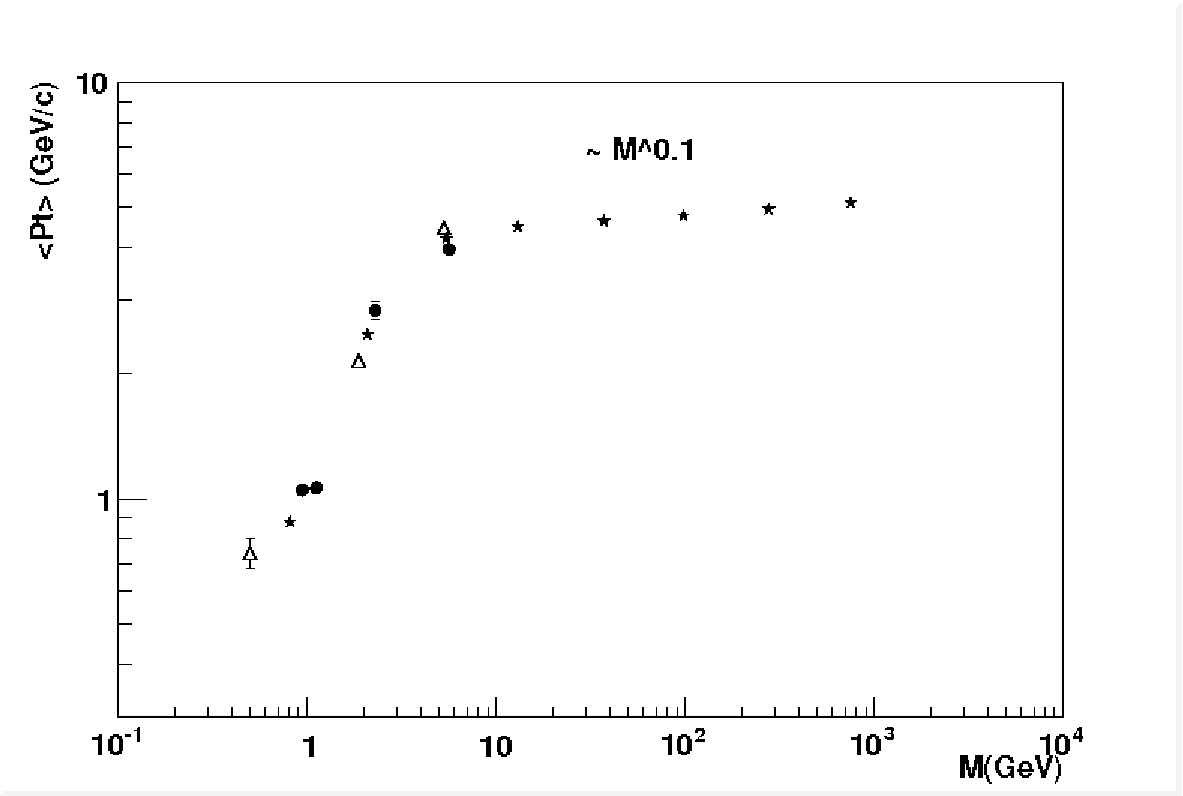}
  \caption{The mass dependence of average momenta of hadrons (mesons - empty triangles, baryons - black circles). The masses of  hypothetical heavy hadron states are shown with the black stars. The mass progression above the beauty hadron masses is expanded with average $p_t$ dependence $<p_t> \propto M^{0.1}$.}
\label{averageVSmass}	
\end{figure}

The dependence of average transverse momenta on the mass of hadrons in the figure~\ref{averageVSmass} shows that $<p_t>$ grows with masses. If we imagine exponentialy symmetric point between meson and baryon masses for each hadron generations, the mass distance between points of one  generation and the other can be estimated with the factor $\delta(lnM)$ = 1. This regularity means the following geometric progression for the masses of hypotetical neutral hadron states: $M_n = 0.25*e^{n-1}$.

The extension of this sequence provides us with the hadron states of the following masses: 0.251, 0.682, 1.85, 5.04, 13.7, 37.2, 101., 275, 748 GeV and so forth. 
The plot is extrapolated above the bottom hadron masses with the average $p_t$ dependence $<p_t> \propto M^{0.1}$ . The hypothetical hadron states may represent heavy multi quark conglomerates that are neutral and having neither electric, nor baryon/lepton charge. This hypothesis allows to predict new particles, which are matching the conditions for the Dark Matter. 
By the way, such "supersymmetric" unification of baryon and meson features have been traditional in the Regge phenomenology and was revisited recently in \cite{brodsky}. Another feature of suggested Baryonium Dark Matter\cite{BDM} is the baryon-antibaryon symmetry that is partially seen in tetraguarks, pentaquarks and hypotetical hexaquarks, which should be build from baryons and antibaryons.

On the other hand, the mass of top quark does not 
match the suggested collection of new hadron states. But, in any case, top quark mass can be released from  multiquark quasi stable state of the lower mass, if we apply the valuable connection energy.   

\section{Conclusions}

The overview of results in transverse momentum distributions of hyperons produced in proton-proton collisions of various energies \cite{hyperon,recent} has revealed a significant change in the slopes of baryon spectra with energy in the region of $p_t$ = 0,3 - 8 GeV/c.

The average $p_t$ values in proton-proton collisions grow steadily as the power of energy, $s^{0.05}$, up to highest LHC energy 7 TeV and above. Therefore, the hadron production   processes at the energies of LHC are not totally random.

The average transverse momentum analysis, through examining the different mass of hadrons, indicates a regularity in the mass gaps between heavy quark baryon-meson generations. This observation gives the possibility for more hadron states with the masses: 0.251, 0.682, 1.85, 5.04, 13.7, 37.2, 101, 275, 748... GeV, which are produced by geometrical progression with the mass factor $\delta(M)$=2.721828. These hadron states can consist of heavy multi quarks and, practically, are proper candidates for the Dark Matter particles, if are suggested almost stable and neutral. The extrapolation for the average transverse momenta of these states seems useful for the futher seach for Baryonium Dark Matter particles \cite{BDM}. 

This study shows that the results of routine phenomenological analysis of experimental data may have far reaching implications for the high energy physics.

\end{document}